\documentclass{article}
\usepackage{amsfonts}
\usepackage{amsmath}

\input{tcilatex}

\begin{document}

\title{Can accelerated expansion of the universe be due to spacetime
vorticity?}
\author{Babur M. Mirza\thanks{%
E-mail: bmmirza2002@yahoo.com} \\
Department of Mathematics, \\
Quaid-i-Azam University, Islamabad 45320, Pakistan}
\maketitle

\begin{abstract}
We present here a general relativistic mechanism for accelerated cosmic
expansion and the Hubble's constant. It is shown that spacetime vorticity
coupled to the magnetic field density in galaxies causes the galaxies to
recede from one another at a rate equal to the Hubble's constant. We
therefore predict an oscillatory universe, with zero curvature, without
assuming violation of Newtonian gravity at large distances or invoking dark
energy/dark matter hypotheses. The value of the Hubble's constant, along
with the scale of expansion, as well as the high isotropy of CMB radiation
are deduced from the model.
\end{abstract}

\section{Introduction}

Accelerated expansion of the universe, as observed, for example, in the
cosmological redshift measurements using type-Ia supernovae (SNe Ia) as
standard candles\ ~\cite{1,2}, implies the need for an expansion energy
effective at least up to the $Mpc$ scale. A number of independent
observations (including the SNe Ia redshift, the Hubble's constant
measurements [3-5], the cosmic microwave background (CMB) [6,7], baryon
acoustic oscillations [8,9], and various cosmological probes [10]), have
measured the contributions of matter and the cosmological constant to the
energy density of the universe, providing an accurate measurement of the
cosmic acceleration [11]. However the amount of energy for this acceleration
implies a hidden or dark form of energy which is approximately three times
of the observed gravitational mass-energy density in the universe.

Within Einstein's general theory of relativity, the observed expansion rate
can be accounted for by including a cosmological constant, whose origin
remains somewhat mysterious. In this context various mechanisms have been
postulated, including new forms of hypothetical particles, or modifications
of the Newtonian-Einsteinian law of gravitation at large distances, among
others. However these theories are specialized in the sense that they fail
to account for other observed features of the universe, such as the high
degree of isotropy in CMB, or even some feature of the expansion, such as
the correct value of the Hubble's constant.

In this work we show that the specific form of the cosmological constant,
hence cosmic acceleration, can be described by spacetime vorticity,
generated by galactic rotations. We show that this vorticity coupled to the
local (galactic) magnetic field provides the requisite push (repulsive
energy) causing the individual galaxies to recede at an accelerated rate. We
are therefore led to an oscillatory universe, where expansion and conversely
contraction rate is determined by local spacetime vorticity, rather than
global geometry (curvature) of the spacetime. Although local perturbation
effects render it difficult to determine nature of the large scale curvature
of the universe, present astronomical observations constrain the spatial
curvature of the universe to be very close to zero [12-16] with less than $%
0.4\%$ experimental error. \ This implies that where as the local spacetime
geometries around galaxies must be determined from the general theory of
relativity, the large scale structure is well described by Friedmann
equations with zero spatial curvature, or equivalently by Newtonian
cosmology in three dimensional Euclidean space. In this spacetime vorticity,
rather than curvature, has the dominant cumulative effect on the accelerated
expansion. We show that locally generated galactic magnetic fields provide
the essential feedback in the cosmic acceleration, which collectively
extends to the $Mpc$ scale.

Large scale magnetic fields have been detected in the intergalactic medium
surrounding galactic clusters [17-21]. These intracluster magnetic fields
can be generated by turbulent gas motions due to massive interactions
between galaxies and the intracluster gas, by galactic winds, and by jets
from active galactic nuclei . Although generated locally within galaxies,
these fields extend much farther in space. Microgauss magnetic fields have
been observed in the intracluster medium of a number of rich clusters, where
the estimates of the regular magnetic field strength for clusters range from 
$0.2-3%
{\mu}%
G$ [22-25] and on supercluster scales, such as for the Coma cluster and the
Abell 1367 cluster (distance scale $40Mpc)$ [26] of the order $0.2-0.6%
{\mu}%
G$. Observations of magnetic fields in the Coma supercluster and in redshift 
$z\simeq 2$ radio galaxies also hint at the existence of widespread
cosmological fields [21]. Magnetic field energy density per galaxy is
therefore generally quite large, comparable to the total matter density in
each galaxy. Also these microgauss fields extend up to many $Mpc$ without
any substantial decaying. This requires a general amplification mechanism,
effective up to very large cosmic distances.

The above observations motivate a magneto-vorticity interpretation for the
energy responsible for the cosmological constant. We shall see below that
this coupling of magnetic energy density and spacetime vorticity gives the
precise form of the expansion as actually observed. At the same time the
coupling of the locally generated magnetic fields and the spacetime
vorticity due to galactic rotation causes the locally prevalent
(intraclusteral) magnetic field energy to extend up to the $Gpc$ scales
without decaying. We show that this leads to the correct (observed) value of
the Hubble's expansion rate, the scale of cosmic expansion, and the
requisite amount of missing energy (three times the observed matter
density), without invoking the dark energy/dark matter hypotheses, or
modifying Newtonian-Einsteinian law of gravity at large distances. We also
show that the discrepancy in the observed value of the Hubble's constant is
due to the inclusion or exclusion of magneto-vorticity effects in the
observational scales involved.

\section{The cosmological model}

Galactic systems, such as galactic clusters, have a residual magnetic field
generated in individual galaxies with non-zero angular momentum. It can be
shown, using general relativistic Maxwell equations in axially symmetric
spacetimes (see [27], and references therein), that for a mass distribution $%
M$, rotating with an angular momentum $J$, the magnetic field coupled to
spacetime vorticity has the form%
\begin{equation}
B=-\frac{B_{0}}{r\sqrt{1-\frac{R_{s}}{r}}}\cos (\varphi -\omega t),\text{ }%
r>R_{s}.  \tag{1}
\end{equation}%
where $R_{s}=2GM/c^{2}$. Using the total energy density equation $\epsilon
_{B}=\int_{0}^{R}4\pi r^{2}B^{2}dr$, up to a radius $R$, we find after
integrating $\xi =\varphi -\omega t$, from $0$ to $\pi $, that the total
energy density due to the magneto-vorticity field (1) is given by, 
\begin{equation}
\epsilon _{B}=4\pi B_{0}^{2}\left[ R+R_{s}\ln \left( R-R_{s}\right) \right] ,%
\text{ }R>R_{s}.  \tag{2}
\end{equation}%
With the magnetic energy density $\epsilon _{B}$ along the radial direction $%
R$ per unit mass, and total energy constant $-k/2$ per unit mass, we
therefore have by energy conservation,%
\begin{equation}
v^{2}-\frac{2GM}{R}+\frac{E_{B}}{4\pi }=-k,  \tag{3}
\end{equation}%
where $v$ is the expansion velocity, related to the Hubble expansion rate $H$
by $v=HR$, with $R$ being the scale factor. This gives the equation%
\begin{equation}
H^{2}-\frac{8\pi G}{3}\rho +\epsilon _{B}=-\frac{k}{a^{2}},  \tag{4}
\end{equation}%
where $H(t)=\dot{R}(t)/R(t)$, and $\epsilon _{B}=E_{B}/4\pi R^{2}$ is the
magnetic energy density, where we have taken $c=1$. Also, for a flat
universe, we take $k=0$. Here $\rho $ is the ordinary matter density.

Acceleration equation for the expansion rate can be derived by
differentiating the energy equation (3). This has two contributions to the
total acceleration. First, velocity due to the gravitational potential
energy is given by,%
\begin{equation}
v_{G}^{2}=-\frac{8\pi }{3}G\rho (t)R(t)^{2}.  \tag{5}
\end{equation}%
Using mass conservation equation $\rho _{0}R_{0}^{3}=\rho R^{3}$ we thus
have for the gravitational acceleration, by differentiating (5),%
\begin{equation}
a_{G}=-\frac{4\pi }{3}G\rho R<0.  \tag{6}
\end{equation}%
Also, the other contribution to acceleration comes from the magnetic energy
density. We see that this energy corresponds to the kinetic energy per unit
mass given by $\epsilon _{B}=v_{B}^{2}/2$, therefore,%
\begin{equation}
v_{B}^{2}=4\pi B_{0}^{2}\left[ R+R_{s}\ln \left( R-R_{s}\right) \right] ,%
\text{ }R>R_{s}.  \tag{7}
\end{equation}%
Since $R>>R_{s}$, we write $\ln \left( R-R_{s}\right) \approx \ln R$. Also
since acceleration $a_{B}=dv_{B}/dt=\left( 1/2\right) dv_{B}^{2}/dR$, we
have for the acceleration corresponding to this velocity 
\begin{equation}
a_{B}=\frac{d^{2}R}{dt^{2}}=4\pi B_{0}^{2}\left[ 1+\frac{2GM}{c^{2}R}\right]
,\text{ }R>0.  \tag{8}
\end{equation}%
Putting for total mass in the disk, $M=2\pi \rho R^{2}$, and using
equipartition of energy $B_{0}^{2}/3$ in the three dimensional space, we get%
\begin{equation}
a_{B}=4\pi B_{0}^{2}+\frac{16\pi ^{2}B_{0}^{2}G}{3c^{2}}\rho R,\text{ }R>0. 
\tag{9}
\end{equation}%
The total acceleration $\left( a_{G}+a_{B},\right) $ in any direction, is
therefore,%
\begin{equation}
\frac{d^{2}R}{dt^{2}}=-\frac{8\pi }{3}G\rho \left[ 1-\frac{2\pi B_{0}^{2}}{%
c^{2}}\right] R+4\pi B_{0}^{2},\text{ }R>0.  \tag{10}
\end{equation}

The acceleration (10) equation has real solution for uniform constant mass
density, given by%
\begin{equation}
R(t)=A\cos \alpha t+\frac{4\pi B_{0}^{2}}{\alpha ^{2}},  \tag{11}
\end{equation}%
where $\alpha =\sqrt{8\pi G\rho \left( 1-2\pi B_{0}^{2}/c^{2}\right) /3}$.
Here $\alpha $ can be real or imaginary corresponding to $\left( 1-2\pi
B_{0}^{2}/c^{2}\right) $ is negative or positive. Putting $R(0)=0$, we have $%
A=-4\pi B_{0}^{2}/\alpha ^{2}$, therefore we have the branches for the
solution%
\begin{align}
R(t)& =\frac{4\pi B_{0}^{2}}{\beta ^{2}}(1-\cosh \beta t),\text{ }4\pi
B_{0}^{2}>c^{2},  \notag \\
& =\frac{4\pi B_{0}^{2}}{\beta ^{2}}(1-\cos \beta t),\text{ }4\pi
B_{0}^{2}<c^{2},  \tag{12}
\end{align}%
where $\beta =\sqrt{8\pi G\rho \left\vert 1-2\pi B_{0}^{2}/c^{2}\right\vert
/3}$. Since for the solution involving $\cosh $, the radial function $R(t)$
is negative, we discard this solution as unphysical. For the remaining part
of the solution we write the magnetic energy density $4\pi B_{0}^{2}$, as
the mean kinetic energy of expansion. Therefore, by virial theorem, we write 
$4\pi B_{0}^{2}=2V^{2}$, where $V^{2}$ is the average energy of expansion
(per unit mass) per $Mpc$. This gives 
\begin{equation}
R(t)=R_{0}(1-\cos \beta t),\text{ }V^{2}<c^{2},  \tag{13}
\end{equation}%
where $\beta =\sqrt{8\pi G\rho \left\vert 1-V^{2}/c^{2}\right\vert /3}$ and $%
R_{0}=V^{2}/\beta ^{2}$.

The acceleration equation (10) also has a solution when $4\pi
B_{0}^{2}=c^{2} $. In this case $\beta =0$ and we have using $R(0)=0$,%
\begin{equation}
R(t)=2\pi B_{0}^{2}t^{2}.  \tag{14}
\end{equation}%
Putting $B_{0}^{2}=c^{2}/2\pi $, by virial theorem, we obtain 
\begin{equation}
R(t)=c^{2}t^{2},\text{ when }\beta =0\text{.}  \tag{15}
\end{equation}%
In this case the universe accelerates as $d^{2}R/dt^{2}=2c^{2}$. This
however implies that, for a non-zero matter density, the average speed of
expansion $V$ must be equal to the speed of light, which is possible only in
the inflation phase of expansion.

\section{Results}

Equation (13) implies a cyclic, oscillating universe. To derive the time and
radius for maximum expansion, we see that the parameter $\beta $ depends on
the matter density $\rho $ of the universe. Observations show that the
matter density at present is approximately $\rho _{0}=\rho (t_{0})\sim
10^{-26}kg/m^{3}$. Also, the observed order of magnitude (up to the $Mpc)$
for the magnetic field is of $\mu G$, we have $V^{2}=4\pi B_{0}^{2}\sim
10^{-11}J/m^{3}$ valid at least up to a $Mpc$, we can therefore neglect $%
V^{2}/c^{2}$ term in $\beta $. This gives for the expansion parameter $\beta
=\sqrt{8\pi G\rho /3}\sim 21\times 10^{-19}/s$, which leads to the time for
maximum expansion $t_{\max }\sim \pi /\beta \sim 10^{19}s$.

Also, equation (13) shows that there is a maximum expansion at $%
2R_{0}=2V^{2}/\beta ^{2}=4\pi B_{0}^{2}/\beta ^{2}$. Since $V^{2}$ is the
energy density per unit mass, per unit distance of $Mpc$, this gives, $%
V^{2}/\beta ^{2}\sim 10^{5}Mpc$. Therefore the maximum radius of expansion
for the universe is approximately $200Gpc$. Figure 1 shows the scale
function for the expansion phase of the universe, corresponding to the
solution (13).

In equation (13) $\beta $ is the expansion rate, related to the Hubble's
parameter $H_{0}=H(t_{0})$ as follows. For the present matter density,
parameter $H_{0}$ is related to $\beta $ by scaling as $\beta
(Mpc/km)\approx H(t_{0})$. Putting the value for $\beta $ (neglecting $%
V^{2}/c^{2}$), and using this scaling we obtain the Hubble parameter at
present as

\begin{equation}
H(t_{0})\sim (21\times 10^{-19}/s)(3\times 10^{22}/10^{3})=63(km/s)Mpc^{-1}.
\tag{16}
\end{equation}%
This value of the Hubbles's parameter however excludes the contribution of
magneto-vorticity term $4\pi B_{0}^{2}$ in the acceleration equation (10).
For a better estimate we therefore use energy equation (3) directly. Putting
back $c^{2}$ factor, we see that for a flat universe $H=H_{0}+H_{B},$ where $%
H_{0}=\sqrt{8\pi G\rho _{0}/3}=H(t_{0})$ and $H_{B}=\sqrt{%
B_{0}^{2}c^{2}/4\pi R^{2}}$. On $Mpc$ scale, $R=1Mpc$ and for average
observed magnetic field $B_{0}\approx 10^{-7}G$. Also, interpreting $H$ as
the total expansion velocity per $Mpc$ with $H_{0}$ and $H_{B}$ as velocity
components per $Mpc$, we have measuring both $H_{0}$ and $H_{B}$ in $km/s$
per $Mpc$, $H_{B}=8.8(km/s)Mpc^{-1}$. This gives for the present rate of
expansion $H\approx 72(km/s)Mpc^{-1}$, which lies in the observed range of
values for the Hubble's parameter (see e.g., Ref. [3-5]). Also, a relevant
parameter to the expansion is the acceleration magnitude. We see from
equation (13) that it is approximately given by 
\begin{equation}
a_{0}=\left\vert \frac{d^{2}R}{dt^{2}}\right\vert =\beta
^{2}R_{0}=H^{2}\left( \frac{V^{2}}{H}\right) =3.9\times
10^{5}km^{2}/s^{2}Mpc^{-1}.  \tag{17}
\end{equation}

In Figure 2 we give the expansion rate $R(t)$ up to the $50Mpc$ of the
observed universe, where the present time is $t=1$ on the horizontal scale.
We notice that the accelerated expansion becomes significant after time
scale of about $0.2\times 10^{17}s$. This value corresponds to the expansion
scale of $Mpc$ for Type Ia supernovae [28,29].

We now deduce the (virial) energy corresponding to this expansion
(acceleration). Since the kinetic energy of expansion $K$ in a flat universe
must be equal to the total potential energy $U$, we have by the energy
balance for local matter distribution%
\begin{equation}
K=U_{B}+U_{G},  \tag{18}
\end{equation}%
which gives on taking average over cluster distributions in the universe, 
\begin{equation}
\left\langle K\right\rangle =\left\langle U_{B}\right\rangle +\left\langle
U_{G}\right\rangle .  \tag{19}
\end{equation}%
By virial theorem, for gravitationally bound systems, $2\left\langle
K\right\rangle +\left\langle U_{G}\right\rangle =0$, which implies that $%
\left\langle U_{G}\right\rangle =-2\left\langle K\right\rangle $. Therefore,%
\begin{equation}
\left\langle U_{B}\right\rangle =3\left\langle K\right\rangle .  \tag{20}
\end{equation}%
Since the dynamical mass is contained in the kinetic energy term, this
implies that the magnetic field contributes $3$ times more mass in the
dynamics, hence expansion, of the universe than the gravitational mass. We
see that this factor comes essentially from virial theorem, which confirms
that the missing dynamical mass must be due to collective (average) effect
of individual (galactic) masses.

Notice that the extent of extragalactic magnetic fields, for a cluster of
galaxies, can be determined by the condition, $\cos (\varphi -\omega t)=0$.
Given a time scale of $6$ billion years for cosmic expansion, and an average
angular momentum $J=mvr=10^{64}m^{2}kg/s$ per galaxy in a cluster
(consisting of a thousand galaxies), we see that for $\varphi =0$, (that is,
along a given direction in space), we have $Jt/r^{3}=\pi /2$. This implies
that $r\sim 10^{27}m\sim 100Gpc$. This again is compatible with the spatial
extent of maximum expansion derived above.

\section{Behavior near singularity}

In the contraction phase, we see that, with sufficient accumulation of mass
density, Schwarzschild singularity $R_{s}$ begins to form. As $R\rightarrow
R_{s}$, we see, however, that the energy density due to the magnetic field
becomes increasingly small, since the logarithmic term in equation (2)
becomes negative. Maxwell equations however imply that under spacetime
reversal, the magnetic field should reverse sign as well. It therefore
follows that magnetic field, hence the magnetic energy density, must be zero
at $R=R_{s}$. However, such reversion implies that the gravitational
attraction must become extremely large as $R$ becomes smaller. Contraction,
therefore, continues up to the radius $R=R_{s}$ with an accelerated rate.
Once cross-over at the Schwarzschild surface occurs, spacetime inversion
must then cause both magnetic and gravitational fields to invert signs as
well. For the interior spacetime we therefore have the magnetic energy
density 
\begin{equation}
\epsilon _{B}=-4\pi B_{0}^{2}\left[ R+R_{s}\ln (R_{s}-R)\right] ,\text{ \ \ }%
R<R_{s},  \tag{21}
\end{equation}%
which now acts as an attractive source of energy, where as the gravitational
field energy $\epsilon _{G}=-GM/R$ is now repulsive for $R<R_{s}$. On
further contraction, as $R\rightarrow 0$, gravitational repulsive energy
exceeds the magnetic attractive energy, and expansion starts in the region $%
R<R_{s}$. For this expansion we have the solution%
\begin{equation}
R(t)=R_{s}-R_{0}(\cosh \beta t-1),\text{ }V^{2}<c^{2}.  \tag{22}
\end{equation}%
Quantum gravity effects, however, imply that the radius $R_{s}$ over which
this sudden expansion occurs, due to very small length scale, lies below the
Planck time $t_{p}$. Since $\epsilon _{B}(R_{s})=0$, this expansion
continues beyond $R=R_{s}$ (Figure 3), due to the gravitational force.

As the solutions posses symmetry, the magnetic energy density in the
neighborhood of cross-over radius $R=$ $R_{s}$ will have a very high degree
of isotropy. Maximum expansion then corresponds to a very high degree of
entropy around cross-over $R=R_{s}$ well into the region $R>R_{s}$. Initial
radiation distribution therefore must be isotropic and of the Planckian
profile due to the thermodynamic equilibrium between matter and radiation.
After the initial gravitational expansion into the region $R>R_{s}$, as $R$
increases, magneto-vorticity effects in spacetime begin to dominate after a
time period of Planck time scale. However, gravity is now attractive,
therefore a shock like condition must result causing matter accumulation,
and vorticity generation.

\section{Conclusions}

To recapitulate we remark that, within the above model of the accelerated
expansion of the universe, local spacetime vorticity and magnetic field
energy generation within galaxies and galactic clusters act as the feedback
mechanism for expansion. Thus contrary to some recent suggestions that
accelerated expansion must imply a violation of the law of conservation of
energy, we see that energy conservation remains strictly valid not only
locally but globally as well. The continued universal acceleration depends
on the energy generation within galaxies, which in turn is determined by
accretion rate in galactic nuclei. Conversion of matter-energy density into
the magnetic field energy under such conditions can only take a finite
amount of time, hence the magnetic field driven acceleration cannot continue
indefinitely for a finite total mass. Since the acceleration $a_{B}\sim
B_{0}^{2}$, where $B_{0}^{2}$ is the magnetic energy density per unit
volume, we see that with the decreasing feedback magnetic field, universal
acceleration after reaching an maximum will gradually decrease. With the
decrease of magnetic energy generation via accretion, a gradual
deacceleration under gravitational attraction is likely to cause cosmic
contraction, as shown above. We therefore have an oscillatory universe,
where magneto-vorticity coupling rather than global spacetime curvature
causes the expansion and contraction phases.

Finally, a very high degree of entropy must have existed at the early stage
of the universe, as inferred from the Planckian shape of the CMB radiation.
This raises the paradox for other cosmological models, since entropy should
decrease closer to the initial singularity (big bang). The above model
implies that this must be so due to the expansion started before the
cross-over $R=R_{s}$. Subsequently, as this expansion (inflation) stops, and
matter formation starts, expansion under spacetime vorticity must now cause
matter entropy to gradually increase with time. As deduced above this
explains the high isotropy and the Planckian profile of the CMB spectrum,
carrying the imprint of this initial inflation over $R>R_{s}$.

\bigskip 

Figure Captions:

Figure 1: Accelerating expansion during the present phase up to the distance
of $50Mpc$, where acceleration effects become significant for redshift
observations of distant Type Ia supernovae.

Figure 2: Scale factor for expansion of the universe. Accelerating expansion
is greatest during the first phase, lasting up to $0.4\times 10^{19}s$,
after which acceleration is almost zero for next $0.2\times 10^{19}s$ and
then decreases, reaching maximum expansion.

Figure 3: Expansion during the first Planck second ($=5.4\times 10^{-44}s$)
of the universe, extending beyond the cross-over radius $R=R_{s}$. Energy
density increase causes matter formation which subsequently slows the
expansion under gravitational attraction.

\bigskip


\begin{thebibliography}{99}
\bibitem{1} Riess, A. G., Filippenko, A. V., Challis, P., Clocchiatti, A.,
Diercks, A., et al. 1998, AJ 116, 1009

\bibitem{2} Perlmutter, S., Aldering, G., Goldhaber, G., Knop, R. A.,
Nugent, P., et al.\ 1999, ApJ, 517, 565

\bibitem{3} Jackson, N. 2015, Living Reviews in Relativity, 18, 2

\bibitem{4} LIGO Scientific Collaboration, et al. 2017, Nature 551(7678),
85\ \ \ 

\bibitem{5} Di Valentino, E, Melchiorri, A. 2018, Physical Review D 97(4),
041301

\bibitem{6} Halverson, N. W., Leitch, E. M., Pryke, C., Kovac, J.,
Carlstrom, J. E., et al. 2002 ApJ, 568, 38 \ 

\bibitem{7} Bennett, C. L., Halpern, M., Hinshaw, G., Jarosik, N., Kogut,
A., et al. 2003 ApJS, 148, 1

\bibitem{8} Seo, H. J., \& Eisenstein, D. J. (2003). ApJ, 598, 720

\bibitem{9} Dawson, K. S., Schlegel, D. J., Ahn, C. P., Anderson, S. F.,
Aubourg, \'{E}., 2012 AJ, 145, 10

\bibitem{10} Ade, P. A., Aghanim, N., Arnaud, M., Ashdown, M., Aumont, J.,
et al. 2016, A\&A, 594, A13\ 

\bibitem{11} Rubin, D., \& Hayden, B. 2016, ApJL, 833, L30\ 

\bibitem{12} Larson, D., Weiland, J. L., Hinshaw, G., \& Bennett, C. L.
2015, ApJ, 801, 9\ 

\bibitem{13} Spergel, D. N., Bean, R., Dor\'{e}, O., Nolta, M. R., Bennett,
C. L., et al. 2007, ApJS, 170, 377

\bibitem{14} Ade, P. A. R., Aghanim, N., Armitage-Caplan, C., Arnaud, M.,
Ashdown, M., et al. 2014 A\&A, 571, A16.11\ 

\bibitem{15} de Bernardis, P., Ade, P. A., Bock, J. J., Bond, J. R.,
Borrill, J., et al. 2000, Nature, 404, 955\ 

\bibitem{16} Rubin, D., Aldering, G., Barbary, K., Boone, K., Chappell, G.,
2015 ApJ, 813, 137

\bibitem{17} Beck, R., \& Wielebinski, R. 2013, in Planets, Stars and
Stellar Systems (Netherlands: Springer), pp. 641-723

\bibitem{18} Kronberg, P. P., Kothes, R., Salter, C. J., \& Perillat, P.
2007, ApJ, 659, 267

\bibitem{19} Beck, R. 2011, AIP Con.Proc. 1381, 117\ 

\bibitem{20} Xu, Y., Kronberg, P. P., Habib, S., \& Dufton, Q. W. 2006, ApJ,
637, 19

\bibitem{21} Widrow, L. M. 2002, Rev. Mod. Phys., 74, 775

\bibitem{22} Taylor, G. B., Barton, E. J., \& Ge, J. 1994, AJ, 107, 1942\ 

\bibitem{23} Kim, K. T., Kronberg, P. P., Dewdney, P. E., \& Landecker, T.
L. 1990, ApJ, 355, 29\ 

\bibitem{24} Kim, K. T., Tribble, P. C., \& Kronberg, P. P. 1991, ApJ, 379,
80

\bibitem{25} Clarke, T. E., Kronberg, P. P., \& B\"{o}hringer, H. 2001,
ApJL, 547, L111\ 

\bibitem{26} Kim, K. T., Kronberg, P. P., Giovannini, G., \& Venturi, T.
1989, Nature, 341, 720

\bibitem{27} Mirza, B. M. 2017, ApJ, 847, 73\ 

\bibitem{28} Riess, A.G., Li, W., Stetson, P.B., Filippenko, A.V., Jha, S.,
Kirshner, R.P., et al. 2005 ApJ, 627, 579

\bibitem{29} Hoffmann, S.L., Macri, L.M., Riess, A.G., Yuan, W., Casertano,
S., et al.\ 2016, ApJ, 830, 10
\end{thebibliography}
\end{document}